\documentclass[superscriptaddress,amsmath,amssymb,showpacs,floatfix,preprintnumbers,final,1p,times,onecolumn]{elsarticle}
\usepackage[latin9,utf8]{inputenc}
\usepackage{float}
\usepackage{textcomp}
\usepackage{amsmath,amssymb,amscd,hyperref}
\usepackage{graphicx}
\makeatletter
\usepackage{url}
\usepackage{grffile}
\usepackage{bbold}
\usepackage{comment}
\usepackage{dirtytalk}
\usepackage{chngcntr}
\usepackage{tabularx}
\usepackage{multirow}
\usepackage{soul}
\usepackage{cleveref}
\DeclareUnicodeCharacter{2212}{-}
\DeclareUnicodeCharacter{0301}{\hspace{-1ex}\'{ }}
\usepackage[dvipsnames]{xcolor}
\usepackage{natbib,hyperref}
\biboptions{sort&compress}

\usepackage{background}
\backgroundsetup{
    scale=10,
    color=red,
    opacity=0.2,
    angle=45,
    position=current page.center,
    vshift=0cm,
    hshift=0cm,
    contents={arXiv Version}
}

\usepackage{dcolumn}
\usepackage{color}
\DeclareGraphicsExtensions{.png .jpg .pdf}
\hypersetup{
     colorlinks = true,
     linkcolor = blue,
     anchorcolor = blue,
     citecolor = blue,
     filecolor = blue,
     urlcolor = blue
     }
\usepackage{braket}
\usepackage{physics}
\usepackage{array}
\usepackage{booktabs}
\usepackage{soul}
\usepackage[marginal]{footmisc}

\makeatother

\begin{document}

\title{Ultrasensitive acoustic graphene plasmons in a graphene-transition metal dichalcogenide heterostructure: strong plasmon-phonon coupling and wavelength sensitivity enhanced by a metal screen}

\author[label1,label2,label3]{I. R. Lavor\fnref{fn1}\corref{cor}}
\affiliation[label1]{
             organization={Instituto Federal de Educação, Ciência e Tecnologia do Maranhão},
             addressline={KM-04, Enseada},
             city={Pinheiro},
             postcode={65200-000},
             state={Maranhão},
             country={Brazil}}
\affiliation[label2]{
             organization={Department of Physics, University of Antwerp},
             addressline={Groenenborgerlaan 171},
             city={Antwerp},
             postcode={2020},
             country={Belgium}}
\affiliation[label3]{
             organization={Departamento de Física, Universidade Federal do Ceará},
             addressline={Caixa Postal
6030, Campus do Pici},
             city={Fortaleza},
             postcode={60455-900},
             country={Brazil}}
\ead{icaro@fisica.ufc.br}

\author[label2]{Z. H. Tao\fnref{fn1}}

\author[label4]{H. M. Dong\corref{cor}}
\affiliation[label4]{
             organization={School of Materials and Physics, China University of Mining and Technology},
             city={Xuzhou},
             postcode={221116},
             country={P. R. China}}
\ead{hmdong@cumt.edu.cn}

\author[label2,label3]{A. Chaves}

\author[label5,label2,label3]{F. M. Peeters}
\affiliation[label5]{
             organization={Nanjing University of Information Science and Technology},
             city={Nanjing},
             postcode={210044},
             country={China}}
             
\author[label2,label6]{M. V. Milo\v{s}evi\'c\corref{cor}}
\affiliation[label6]{
             organization={Instituto de Física, Universidade Federal de Mato Grosso},
             city={Cuiabá},
             postcode={78060-900},
             state={Mato Grosso},
             country={Brazil}}
\ead{milorad.milosevic@uantwerpen.be}

\fntext[fn1]{I. R. Lavor and Z. H. Tao contributed equally to this work.}
\cortext[cor]{Corresponding author}

\date{\today }

\begin{abstract}  
Acoustic plasmons in graphene exhibit strong confinement induced by a proximate metal surface and hybridize with phonons of transition metal dichalcogenides (TMDs) when these materials are combined in a van der Waals heterostructure, thus forming screened graphene plasmon-phonon polaritons (SGPPPs), a type of acoustic mode. While SGPPPs are shown to be very sensitive to the dielectric properties of the environment, enhancing the SGPPPs coupling strength in realistic heterostructures is still challenging. Here we employ the quantum electrostatic heterostructure model, which builds upon the density functional theory calculations for monolayers, to show that the use of a metal as a substrate for graphene-TMD heterostructures (i) vigorously enhances the coupling strength between acoustic plasmons and the TMD phonons, and (ii) markedly improves the sensitivity of the plasmon wavelength on the structural details of the host platform in real space, thus allowing one to use the effect of environmental screening on acoustic plasmons to probe the structure and composition of a van der Waals heterostructure down to the monolayer resolution. 
\end{abstract}


\maketitle

\section{Introduction}\label{Sec. Intro}

Graphene, a two-dimensional (2D) material made of carbon~\cite{Novoselov2004}, has been extensively investigated over the past decades in various fields of photonics and opto-electronics engineering~\cite{bonaccorso2010graphene}, for use in photovoltaic cells~\cite{messina2013graphene}, photodetectors~\cite{wang2013high,liu2014graphene}, terahertz (THz) devices~\cite{vicarelli2012graphene,zhang2022strongly}, and light-emitting devices~\cite{han2012extremely}, and many other examples. Furthermore, 2D transition metal dichalcogenides (TMDs) of MX$_2$ form, where $M$ is a transition metal and $X$ is a chalcogen element (e.g., MoS$_2$, MoSe$_2$, WS$_2$, WSe$_2$), have also attracted significant interest for their exceptional opto-electronic characteristics~\citep{Geim2013,Low2014,Low2016,Ranieri2014,Jariwala2014,Zhang2015,Ju2011,Chen2012,Fiori2014,Mak2016}. When graphene is combined with TMDs into van der Waals heterostructures (vdWhs), which can be done either by vertically stacking layers of different TMDs,~\citep{Geim2013,Novoselov2016,Wang2012,Jariwala2014,Zhang2015,Mak2016,gong2014vertical,Manzeli2017} or by arranging them side by side to form lateral vdWhs~\cite{ozcelik2016band,sahoo2018one,duan2014lateral,gong2014vertical,gong2015two,huang2014lateral,Manzeli2017}, it becomes possible to engineer a range of multi-layered artificial materials, each exhibiting unique properties that can be tailored practically at will.

In the field of plasmonics, the excitation of electrons in graphene by light in the THz to mid-infrared range is known to lead to optical graphene plasmon-polaritons~\citep{Low2014,Goncalves2015,Ju2011,Fei2012}, i.e. collective oscillations of the two-dimensional electron liquid in graphene~\citep{GabrieleGiuliani2008,Maier2007}, commonly referred to as Dirac plasmons~\citep{Grigorenko2012}. In this context, significant progress has been made towards the development of new plasmon-based devices using 2D nanomaterials and graphene, driven by their vast promising applications in sensors~\cite{ogawa2020graphene}, light-emitting devices~\cite{fusella2020plasmonic}, nano-optical systems~\cite{liu2018switch,sun2016optical}, spectroscopy~\cite{eberlein2008plasmon}, molecular fingerprints~\cite{voronin2020nanofocusing}, and radiative heat transfer~\cite{lu2022enhanced}. Furthermore, the possibility to excite plasmons in graphene combined with different 2H-TMDs, where 2H refers to the hexagonal symmetry~\cite{Wang2012}, thus forming different vdWhs, opens up innovative avenues for developing low-dimensional technologies~\cite{zhang2022plasmonic}. This potential stems from the high sensitivity of graphene plasmons to modifications in their surrounding environment~\cite{tao2021tailoring, lavor2021tunable}, strong electric field confinement, and low losses~\cite{Fei2012}. Furthermore, graphene plasmons exhibit tunability through the modulation of the Fermi level, e.g. via back-gating techniques~\cite{williams2011gate}, thereby adjusting the density of charge carriers. All these properties make graphene an exceptionally suitable base for the development of a wide array of optoelectronic devices.

\begin{figure}[!t]
\centering{}\includegraphics[width=1\columnwidth]{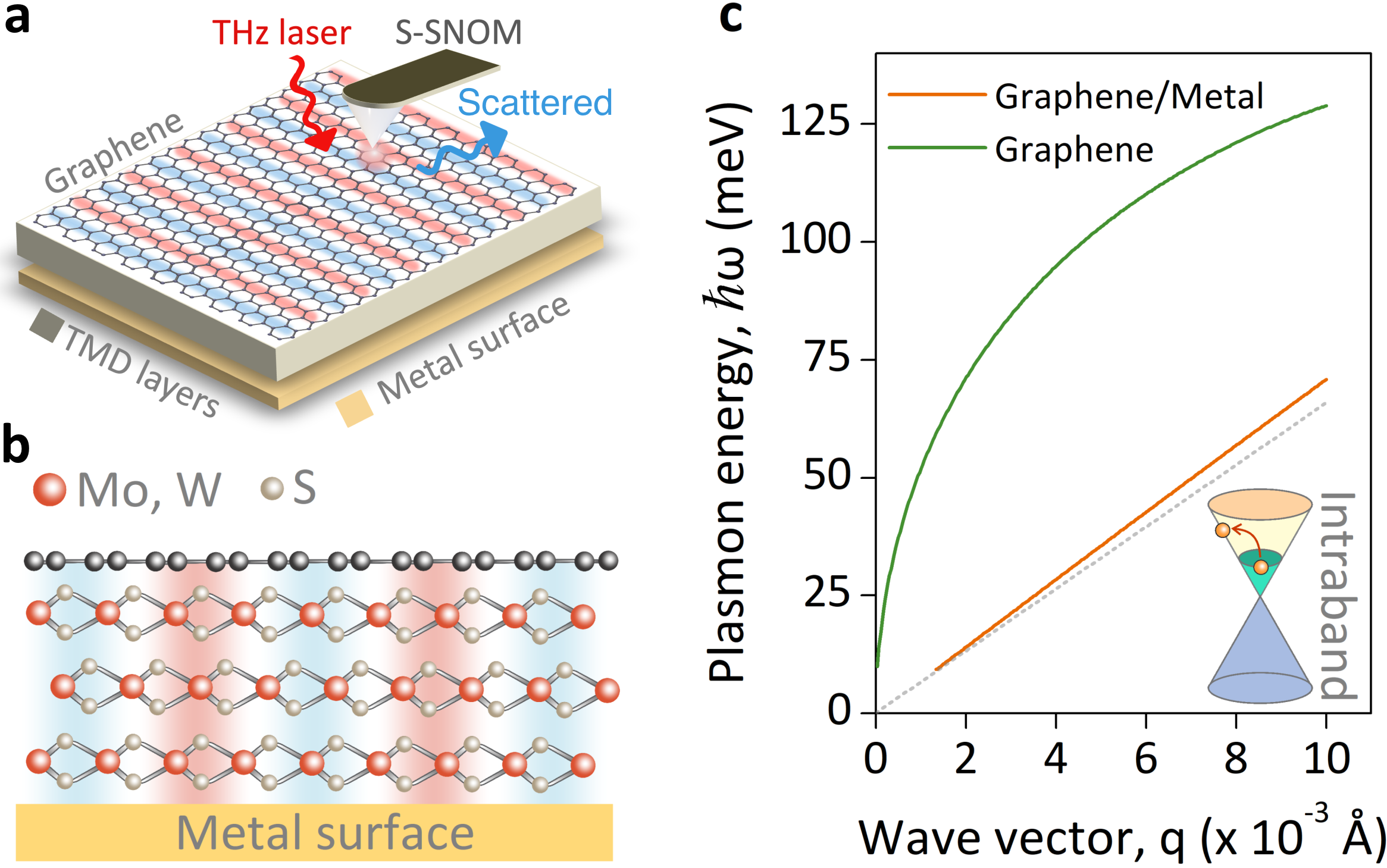}
\caption{(Color online) (a) Schematic illustration of the scatter-type scanning near-field optical microscopy setup (s-SNOM) and an acoustic Dirac plasmon wave in a van der Waals heterostructure.   The heterostructure is composed of a monolayer graphene (G) on top of $N$-layer of transition metal dichalcogenides (TMDs) with a metal surface (M) underneath. The TMDs have the $N$-MX$_2$ form, where $N$ represents the number of layers and $\text{M} = \text{Mo, W}$; $\text{X} = \text{S}$. (b) Lateral view of a G/3-MX$_2$/M vdWhs. The presence of the metal surface gives rise to screened graphene plasmon-phonon polaritons (SGPPPs), the highly confined acoustic plasmon modes. (c) Dispersion of optical Dirac plasmons in a free-standing graphene monolayer, with a well-known $\sqrt{q}$ dependence (solid green curve), and acoustic Dirac plasmons due to the presence of a nearby metal surface (solid orange curve) with a linear dependence on the wave-vector $q$. Both results are for a Fermi energy set at $E_F = 100~\text{meV}$. The dotted gray line delimits the intraband transition.}
\label{Fig: dispersion and structure}
\end{figure}

Although the coupling of surface plasmon polaritons with matter and/or other quasi-particles (phonons and excitons, for example) has been investigated in different systems, such as three-dimensional gold nanoparticles~\cite{Mueller2020}, thermal emitter atomic system~\cite{huang2023strong}, graphene-based vdWhs~\cite{lavor2021tunable,tao2023ultrastrong}, plasmon-exciton polaritons in 2D semiconductors/metal~\cite{gonccalves2018plasmon} and molecular sensing~\cite{voronin2020nanofocusing}, a comprehensive study examining how a proximate metallic surface influences the interaction between graphene plasmons and TMD phonons in a vdWhs, on a layer-by-layer basis, remains to be conducted. Therefore, our study specifically addresses this gap, building on recent advancements such as the broadband enhancement of Cherenkov radiation using dispersionless plasmons in graphene/MX$_2$/metal configuration, as reported in Ref.~\cite{hu2022broadband}, to provide novel insights into the plasmon-phonon interactions in such vdWhs on metallic substrates. To do so, in this work, we investigate a vdWhs composed of multiple TMDs layers covered by monolayer graphene on top, as a plasmonic material, and placed over a metal substrate, as illustrated in Fig.~\ref{Fig: dispersion and structure}(a,b), which allows us to explore electromagnetically    screened graphene plasmons (SGPs)~\cite{alcaraz2018probing,Alonso-Gonzalez2016,menabde2021real,lee2019graphene}. These SGPs are represented by acoustic modes that are highly confined in the out-of-plane direction, thus boosting the coupling strength with out-of-plane TMDs phonon modes. Also, in the THz regime, SGPs do not experience Landau damping. The presence of the metal results in an efficient screening of the graphene plasmons, confining the SGPs into the narrow space between the graphene and the metal without reducing their lifetime~\cite{alcaraz2018probing}. This results in a highly efficient plasmonic nanocavity with embedded active hybrid plasmon-phonon coupling due to the phonons in TMDs~\cite{kittel2018introduction,Goncalves2015}. This hybrid excitation arises when phonons in the TMDs are coupled to the acoustic electron oscillations in graphene~\cite{Goncalves2015}, giving rise to the screened graphene plasmon-phonon polaritons (SGPPPs). Experimentally, they can be both excited and detected using scatter-type scanning near-field optical microscopy (s-SNOM) techniques~\cite{Lundeberg2017,Lundeberg2016,Alonso-Gonzalez2016}.

This scenario allows us to investigate the coupling strength between graphene plasmons and TMDs phonons in such a polaritonic platform and explore the possibility of enhancing the plasmon wavelength resolution in the real space~\cite{lavor2020probing,lavor2021tunable}. To do so, we employ the quantum electrostatic heterostructure model (QEH)~\cite{Andersen2015}, which integrates the random phase approximation (RPA) with density functional theory (DFT), for evaluating the plasmon-phonon dispersion. This DFT-based model enables us to explore how the properties of plasmons are influenced by the number of TMD layers in the heterostructure and the graphene Fermi energy. Additionally, the use of QEH also allows one to properly account for substrate-induced effects~\cite{Gjerding2020}, such as the metal substrate screening, for example. Interestingly, integrating a metal substrate not only gives rise to SGPs, as expected, but also substantially increases the coupling strength between SGPs and TMD phonons, amplifying it by an order of magnitude, leading to the ultra-strong coupling regime. Taking advantage of the sensitiveness of the plasmon-phonon coupling to the number and composition of TMD layers in the vdWhs, enhanced by the metal substrate, we demonstrate the possibility of using SGPs to probe the structure and composition of the TMD stack below the graphene layer at the ultimate resolution limit of a single monolayer.

The paper is organized as follows. In section \ref{sec:THEORETICAL APPROACH}, we introduce the DFT-based QEH model for simulating the hybrid plasmon-phonon polaritons dispersions. 
In section \ref{sec: acoustic graphene plasmon}, we show the electric field distribution induced by the acoustic plasmon within a heterostructure comprising graphene with a nearby metal substrate. This analysis encompasses varying the separation of graphene from the metal substrate and different excitation frequencies.
Subsequently, we demonstrate in section \ref{sec. lossfunc} how the metal surface influences the loss function.
In section \ref{sec. enhancing}, we reveal the strongly enhanced plasmon-phonon coupling strength attributed to the presence of a nearby metal surface. The influence of doping on the SGPPPs spectra is discussed as well.
Finally, in section \ref{sec: probing} we show the application of SGPPPs in assessing vdWhs composition with ultimate resolution.
Our findings and conclusions are summarized in section \ref{sec:conclusion}.

\section{Methods}
\subsection{Computational model for plasmon-phonon dispersion}
\label{sec:THEORETICAL APPROACH}

The QEH model is employed to calculate the plasmonic properties of the vdWhs~\cite{Andersen2015}.  It incorporates the full electronic response of the vdWhs where it combines RPA and first-principles DFT with the Perdew–Burke–Ernzerhof exchange energy formulation~\cite{Andersen2015}. Within this approach, the QEH enables one to take into account doping contribution, screening response from optical phonons of TMDs in the IR-regime and the bulk substrate~\cite{gjerding2021recent}. Its major advantage is that it uses the dielectric building blocks (DBBs) of 2D materials that were previously obtained with thorough \textit{ab initio} calculations, thus making the QEH a fast and precise open computational model (since all DBBs are available from the Computational 2D Materials Database project (C2DB)~\cite{gjerding2021recent,Link1}). This integration facilitates an in-depth investigation of the interaction between SGPs and the active IR TMD phonon modes in horizontally aligned vdWhs.

Not less important, the QEH method has been demonstrated quite efficient even in complex scenarios involving a large number of vdWhs layers, yielding highly satisfactory results when compared with s-SNOM-based experiments for graphene on top of hexagonal boron nitride (h-BN) or graphene encapsulated by hBN, as demonstrated e.g. in Ref.~\cite{lavor2020probing}. The QEH model has also been successfully used to support experimental results through calculations of electron energy loss spectroscopy, e.g. of multilayer MoS$_2$~\cite{nerl2017probing}.

Next, we provide a comprehensive overview of the methodology used in this paper. For a more detailed description of the QEH method, including information on the DFT methods and $k$-point optimization related to the dielectric building blocks (DBBs) of the C2DB, we refer the reader to Refs. \cite{Andersen2015, Gjerding2020} and their Supplementary Materials.

To investigate the loss function and plasmon dispersion in vdWhs, the QEH employs the density-density response functions \(\chi_{i}(z,z',\mathbf{q}_{\parallel},\omega)\) from each distinct layer (indexed as \(i\)-th) within the DBBs of the C2DB database. Therefore, the QEH combines all these response functions of individual layers, or DBBs, through the long-range Coulomb interactions. To do so, the QEH solves a Dyson-like equation, resulting in the comprehensive density-density response function for the entire vdWh, given by:
\begin{equation}
\chi_{i\alpha,j\beta}=\chi_{i\alpha}\delta_{i\alpha,j\beta}+\chi_{i\alpha}\sum_{k\neq i,\gamma}V_{i\alpha,k\gamma}\chi_{k\gamma,j\beta}.
    \label{Eq: chi QEH}
\end{equation}
In Eq.~(\ref{Eq: chi QEH}), $\alpha=0,1$ represents the monopole and dipole components, respectively. Note that the variables $\mathbf{q}_{\parallel}$ and $\omega$ are implied but not explicitly shown, for brevity. The Coulomb interaction matrices are defined as $V_{i\alpha,k\gamma}(\mathbf{q}_{\parallel})=\int\rho_{i\alpha}(z,\mathbf{q}_{\parallel})\Phi_{k\gamma}(z,\mathbf{q}_{\parallel})dz$, where $\Phi_{k\gamma}(z,\mathbf{q}_{\parallel})$ represents the potential which arises due to  the density profile $\rho_{k\gamma}(z,\mathbf{q}_{\parallel})$, calculated by solving a 1D Poisson equation with open boundary conditions. Employing this approach enables the determination of the vdWhs' inverse tensorial dielectric function, defined as:
\begin{equation}
\epsilon_{i\alpha,j\beta}^{-1}(\mathbf{q}_{\parallel},\omega)\!=\!\delta_{i\alpha,j\beta}+\sum_{k\gamma}V_{i\alpha,j\beta}(\mathbf{q}_{\parallel})\chi_{k\gamma,j\beta}(\mathbf{q}_{\parallel},\omega)~.
\label{Eq. QEH dielectric function}
\end{equation}
In Eq.~(\ref{Eq. QEH dielectric function}), there are three main elements: (i) $\delta_{i\alpha,j\beta}$ ensures that the matrix retains its identity property when there is no interaction between different layers, for the same layer; (ii) $V_{i,\alpha,j\beta}$ represents the Coulomb interaction between different layers and how these interactions affect (iii) the overall dielectric response of the heterostructure, $\chi_{k,\gamma,j\beta}$.

Therefore, the loss function can be determined via
\begin{equation}
L\left(\mathbf{q}_{\parallel},\omega\right)=-
\text{Im}\left[\text{Tr}\left(\frac{1}{\epsilon\left(\mathbf{q}_{\parallel},\omega\right)}\right)\right],
\label{equation_lossfunction}
\end{equation}
and the collective modes, such as the plasmon dispersion, are obtained as the maxima of the loss function.

An important aspect in vdWhs is the interaction between TMDs phonons and SGPs when those are combined. These hybrid modes modify the plasmon dispersion within the \((q,\omega)\)-plane when the SGPs match with the IR-active phonon frequency of the TMD, giving rise to SGPPPs. To accurately capture this aspect, the QEH takes into account the vibrational phonon modes of the constituent 2D materials, obtained previously by \textit{ab initio} calculations at \(\Gamma\)-point, and their Born effective charges~\cite{Gjerding2020}. The Born effective charges, defined as tensors, that represent the changes in 2D polarization density \(P_i\) arising from atomic displacement, are defined as:
\begin{equation}
    Z_{i, aj} = \frac{\text{A}_\text{cell}}{e} \left.\frac{\partial P_i}{\partial u_{\text{aj}}}\right|_{E=0}~.
\end{equation}
Here, \(\text{A}_\text{cell}\) denotes the in-plane area of the 2D layer, \(a\) refers to an atom, and \(i\),\(j\) represent Cartesian coordinates~\cite{Gjerding2020}.

Regarding the phonons, in 2H-TMDs, four types of phonon modes are observable at the $\Gamma$-point within the infrared (IR) spectrum, but only two of them are capable of interacting with graphene plasmons through long-range Fr\"ohlich interactions near their respective phonon frequencies~\cite{Zhang2015a,Ataca2011}. These are the in-plane $\text{E}^{\prime}$ mode, active in both IR and Raman (R) spectroscopies, and the out-of-plane $\text{A}^{\prime\prime}_2$ mode, which is IR-active.  Table \ref{Tab: MX2-AppendixA} summarizes these phonon frequencies and modes for monolayer MoS$_2$ and WS$_2$. It is important to mention that for TMD layers stacked in even numbers, the $\text{E}^{\prime}$ and $\text{A}^{\prime\prime}_2$ modes from a single-layer MX$_2$ (1L-MX$2$) split into E$_{{u}}$ (IR-active) and E$^1_g$ (Raman-active) modes for the former, and A$_{{2u}}$ (IR-active) and A$^1_{1g}$ (Raman-active) modes for the latter. In the bulk form of MX$_2$, the out-of-plane $\text{A}^{\prime\prime}_2$ mode further splits into A$_{2u}$ (IR-active) and B$^{1}_{2g}$ (optically inactive or silent) modes. For a detailed exploration of phonon modes at the $\Gamma$-point in multi-layered 2H-TMDs, we refer to Refs.~\cite{Zhang2015,Ataca2011}.
\begin{table}[!t]
\centering{}

\begin{tabular}{p{0.6cm} p{0.25cm} p{1.35cm} p{1.35cm} p{1.35cm} p{1.35cm}}
\hline\hline

&&   \multicolumn{4}{c}{Phonon frequencies (meV)}  \\ 
\hline
 && 1 (E$^{\prime\prime}$) & 2 (E$^\prime$) & 3 (A$_{1}^\prime$) & 4 (A$_{2}^{\prime\prime}$)\\
\hline
MoS$_{2}$ && 34.19 & 46.35 & 47.59 & 56.16\\
\hline
WS$_{2}$ && 35.56 & 42.85 & 50.12 & 51.00\\
\hline\hline
\end{tabular}
\label{Tab: MX2-AppendixA}

\caption{Optical phonon frequencies for free-standing monolayers $\text{MoS}_{2}$ and $\text{WS}_{2}$, considered in the QEH calculations. The vibrational optical phonon modes of the monolayers are represented by $\text{E}^{\prime\prime}$ (R), $\text{E}^{\prime}$ (IR and R), $\text{A}^{\prime}_1$ (R) and $\text{A}^{\prime\prime}_2$ (IR), where IR (R) means that the mode is active for infrared (Raman) excitations~ \cite{Zhang2015a,Zhao2013,MolinaSanchez2011,Peng2016,Berkdemir2013,Sengupta2015}.}

\end{table}

Another fundamental variable in 2D systems is the lattice polarizability. In the QEH model, at the optical limit ($\mathbf{q}=0$), the polarizability is given by~\cite{Gjerding2020}:
\begin{equation}
    \alpha_{ij}^{\text{lat}}(\omega)\hspace{-0.8mm} =\hspace{-0.8mm} \frac{e^2}{\text{A}_\text{cell}} \hspace{-1mm} \sum_{ak,bl}  Z_{i, ak}[(\mathbf{C} - \mathbf{M}(\omega^2 -i\gamma \omega))^{-1}]_{ak,bl}  Z_{j, bl} ~,
    \label{eq: lattice polarizability}
\end{equation}
where $\mathbf{C}$ represents the force constant matrix at the optical limit, $\mathbf{M}$ is the matrix of atomic masses, and $\gamma$ is a relaxation rate. For a detailed derivation of Eq.~(\ref{eq: lattice polarizability}) we refer to the Supporting Information of Ref.~\cite{Gjerding2020}).

Finally, incorporating both electron and phonon contributions, the total monopole and dipole components of the DBBs for $i$-th layer are~\cite{Gjerding2020}:
\begin{subequations}
\begin{equation}
\chi_{i0}^{\text{total}}\left(\mathbf{q}_{\parallel},\omega\right)=\chi_{i0}^{\text{el}}\left(\mathbf{q}_{\parallel},\omega\right)-\mathbf{q}_{\parallel}^{2}\alpha_{\parallel}^{\text{lat}}\left(\omega\right),
\label{Eq: DBBsa}
\end{equation}
and
\begin{equation}
\chi_{i1}^{\text{total}}\left(\mathbf{q}_{\parallel},\omega\right)=\chi_{i1}^{\text{el}}\left(\mathbf{q}_{\parallel},\omega\right)-\alpha_{zz}^{\text{lat}}\left(\omega\right)~,
\label{Eq: DBBsb}
\end{equation}
\end{subequations}
where $\alpha_{\parallel}^{\text{lat}}$ denotes the $2\times2$ in-plane submatrix of $\alpha^{\text{lat}}$. 

\section{Results and discussion}
\subsection{Acoustic graphene surface plasmons adjacent to a metal}\label{sec: acoustic graphene plasmon}

Let us first provide a concise overview of the characteristics of acoustic graphene plasmons in a vdWhs, in the presence of a metal, as illustrated in Fig.~\ref{Fig: dispersion and structure}(a,b). Figure~\ref{Fig: dispersion and structure}(c) shows how the presence of the metal modifies the plasmon dispersion, as obtained via QEH method, by comparing the plasmon dispersion as a function of the in-plane wave vector $q$ for a freestanding graphene monolayer (solid green curve) with that of graphene placed directly above a metal substrate (solid orange curve), both at a Fermi energy of $E_F = 0.1~\text{eV}$. For this analysis, and throughout this paper, considering the QEH simulations, the distance between the substrate and the layer immediately above it is set to $3~\text{\AA}$. The dotted gray line in Fig.~\ref{Fig: dispersion and structure}(c) marks the boundary of intraband transitions. Below this line, the group velocity of the plasmons in graphene falls below that of the electrons, leading to the damping of plasmon waves. The presence of a metal surface near the graphene monolayer fundamentally changes the plasmon dispersion~\cite{gonccalves2018plasmon,alcaraz2018probing,Alonso-Gonzalez2016,menabde2021real,lee2019graphene}: in the absence of a metal, it scales with $\sqrt{q}$, while the presence of a metal substrate reshapes it into an acoustic plasmon mode with $\hbar\omega\propto v_g q$ instead, where $v_g$ is the SGPs group velocity, as seen in Fig.~\ref{Fig: dispersion and structure}(c). Considering a vdWhs with a few-layer TMD slab between the graphene layer and the metal substrate, for sufficiently large thickness of the TMD, the confinement of the SGPs diminishes and the dispersion of SGPs will scale similarly to monolayer graphene plasmons. In other words, pushing the metal surface further from the graphene monolayer reduces the electron interaction with its mirror charge, thus leading to weaker screening by the metal. Conversely, when the distance $d$ between the graphene layer and the metal substrate (approximately the thickness of the TMD slab in the vdWhs) is comparable to the confinement length of the graphene plasmons, defined as $2\pi/q$, the system starts to exhibit the characteristics of a nanoresonator~\cite{chen2017acoustic}, as a result of the interaction of the graphene plasmons with their mirror image in the metal. Consequently, SGPs emerge, where this plasmon mode resembles the antisymmetric plasmon mode found in a double-layer graphene configuration with twice the distance $d$ between the carbon layers~\cite{ferreira2020quantization,tao2023ultrastrong}.

To provide a better insight into the behavior of the system, we perform numerical finite-element method simulations of the plasmon-induced electric field distribution for a heterostructure made by monolayer graphene on top of a TMD slab with a thickness of $10~$nm, which corresponds to approximately 15 layers of MoS$_2$, with a metal substrate. To do so, we considered graphene as a conducting interface/boundary that sustains the confined electronic-magnetic field mode~\cite{Goncalves2015}, with a Drude surface conductivity defined as~\cite{gusynin2006}
\begin{equation}
    \sigma(\omega) = \frac{e^2E_F}{\pi\hbar^2}\frac{i}{\omega+i\tau^{-1}}.
\label{Eq: sigmaomega}
\end{equation}
In Eq. \eqref{Eq: sigmaomega}, $\tau=0.1~$ps is the lifetime for electrons, with its value set to reproduce a realistic scenario commonly used in optical experiments~\cite{garcia2014graphene}. For the sake of simplicity, we define the dielectric constant in the region in-between graphene and the metal surface as $\epsilon=4 \epsilon_0$. The metal is described by the following Drude model
\begin{equation}
    \epsilon_{M}(\omega) = 1-\frac{\omega_{M}^2}{\omega(\omega+i\gamma_M)},
\end{equation}
where $\omega_{M}$ and $\gamma_M$ are the intrinsic plasmon frequency of the metal and the damping constant, correspondingly. Considering the metal as a perfect electric conductor, its dielectric constant $\epsilon_M\rightarrow-\infty$, reflecting that the intrinsic plasmon frequency of the metal, $\omega_{M}$, is significantly greater than the excitation frequency, $\omega$, i.e. $\omega_{M} \gg \omega$.

\begin{figure}[!t]
\centering{}\includegraphics[width=0.95\columnwidth]{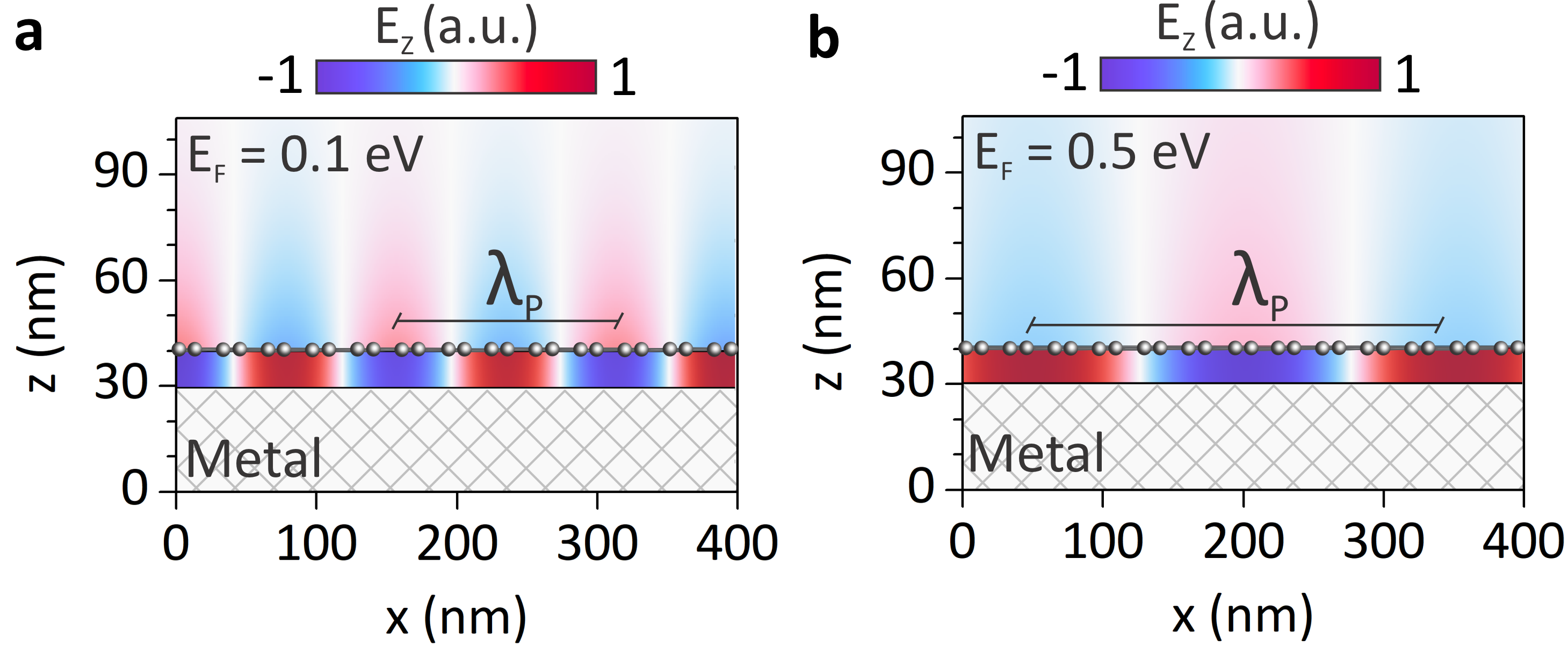}
\caption{(Color online) Color map showing the normalized out-of-plane component of the electric field $E_z$ in a graphene/TMD slab/metal system, at Fermi energies (a) $E_F=0.1~$eV and (b) $E_F=0.5~$eV. The system consists of graphene on top of a slab with a thickness of $10~$nm (equivalent e.g. to $\approx$ 15 MoS$_2$ layers) and dielectric constant $\epsilon=4$. Below the TMD slab is a metal, represented by the hatched area, where the electric field lines are screened. The plasmon wavelength is represented by $\lambda_\text{P}$.}
\label{Fig: electric field}
\end{figure}

Under normal-incidence of an electromagnetic plane-wave $\mathbf{E} = E_0e^{i(k_0z−\omega t)} \hat{x}$ in this system, the resulting $x-$ and $z-$components of the electric field are plotted in Fig. \ref{Fig: electric field} for an excitation frequency ($\omega$) of $1~$THz. To further increase the confinement of SGPs, one can also tune the graphene Fermi energy. The highly vertical electric field polarization of such confined plasmons is expected to strengthen the coupling to out-of-plane TMD phonon modes, since both the plasmon and the phonon modes would oscillate in the same direction. Figures \ref{Fig: electric field}(a) and \ref{Fig: electric field}(b) show the calculated $z-$component of the electric field for $E_F=0.1~$eV and $E_F=0.5~$eV respectively. The in-plane $x$-component has been omitted, being very weak.

Increasing the Fermi energy leads to an increase in the SGPs wavelength ($\lambda_P$), as shown in Figs.~\ref{Fig: electric field}(a) and \ref{Fig: electric field}(c).  Conversely, decreasing the distance between graphene and the metal, increases the confinement but significantly diminishes $\lambda_P$, since the hybridization between the SGPs and their mirror charge intensifies, resulting from the amplified resonance due to the metal screening~\cite{voronin2020nanofocusing,lee2019graphene,Alonso-Gonzalez2016}. In this case, tuning the Fermi energy plays a crucial role in enhancing the sensitivity of SGPs.

\subsection{Effects of the metal surface on the loss function}\label{sec. lossfunc}

\begin{figure}[!t]
\centering{}\includegraphics[width=1\columnwidth]{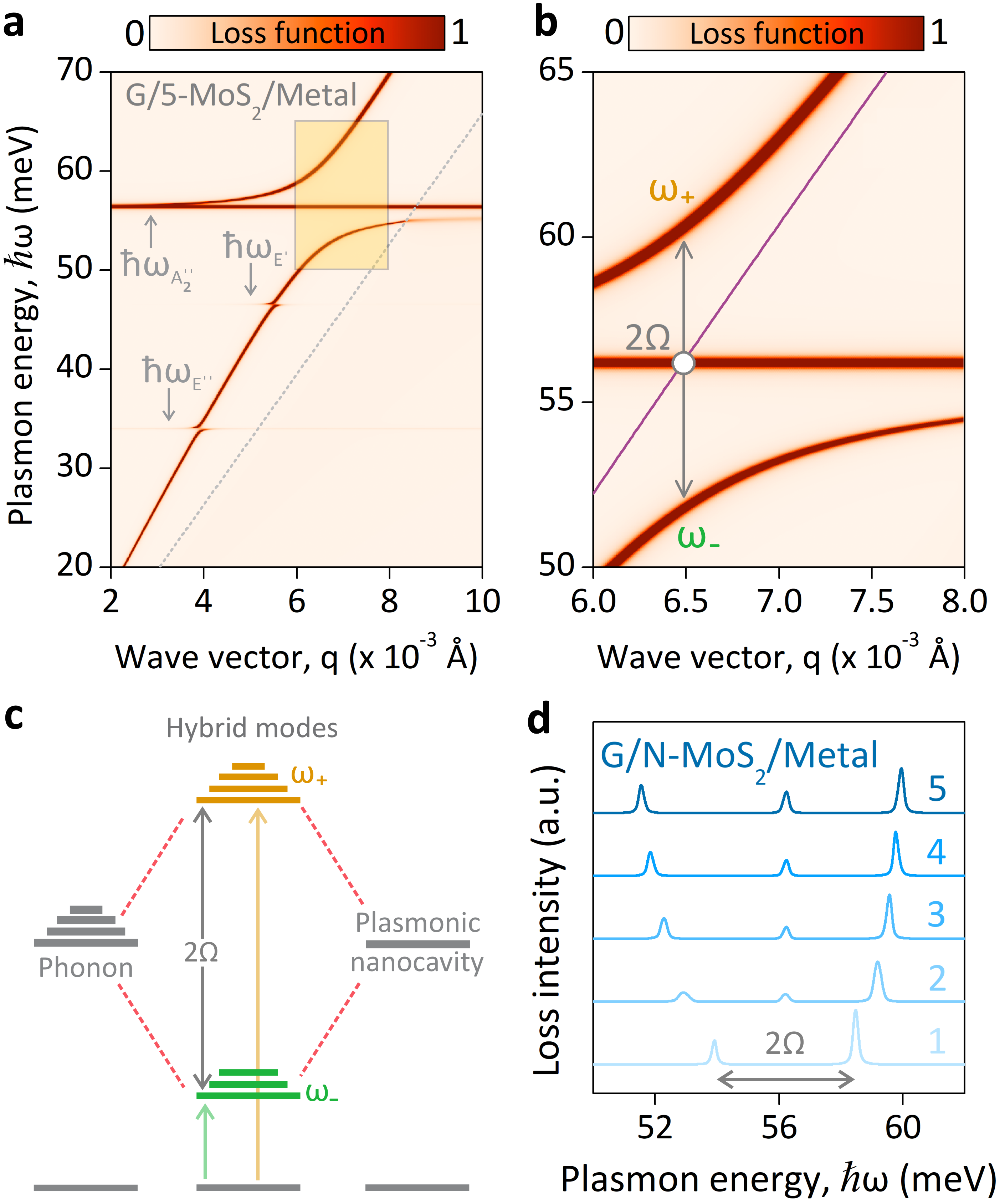}
\caption{(Color online) (a) Loss function of the surface plasmon-phonon polariton dispersion for a G/5-MoS$_2$/M vdWhs with $E_F=100~$meV. The dotted gray line delimits the intraband transitions. (b) Magnification of the yellow area in (a) to emphasize the coupling between the acoustic graphene plasmon and the A$_2^{\prime\prime}$ out-of-plane MoS$_2$ phonon modes. $\Omega$ represents the acoustic graphene plasmon-phonon coupling energy, calculated at the minimum energy splitting (MES), defined as $2\Omega=\omega_{+} - \omega_{-}$. For reference, the purple solid line represents the plasmon dispersion without the contribution of TMD phonons ($\Omega = 0$). (c) Schematic illustration of a two-level system that can represent the coupling of the out-of-plane TMD phonon mode and the acoustic graphene plasmon modes (plasmonic nanocavity). The green (orange) bars represent the bottom (upper) hybrid modes. (d) Loss intensity for G/$N$-MoS$_2$/M, for N from 1 to 5, where N represents the number of MoS$_2$ layers, calculated at the MES (curves are displaced vertically).}
\label{Fig: coupling and loss intensity}
\end{figure}

In Fig.~\ref{Fig: coupling and loss intensity}(a), we present the loss function of SGPs modes, obtained by the QEH calculations, for a G/5-MoS$_2$/M vdWhs, where M represents the metal, with graphene Fermi energy $E_F=0.1~$eV. The region where the group velocity of SGPs is slower than the Fermi velocity, i.e. the intraband region~\cite{Hwang2007}, associated with electronic transitions between levels within the same energy band (see inset in Fig~\ref{Fig: coupling and loss intensity}(a)) is delimited by the dotted gray line. The SGPs exhibit  linear dispersion before hybridizing with the IR-active MoS$_2$ phonon modes, but only with significant coupling to the out-of-plane A$_2^{\prime\prime}$ phonon mode, since both oscillate in the same direction. This hybridization, highlighted in Fig.~\ref{Fig: coupling and loss intensity}(b), leads to SGPPPs and will be the focus of our further investigation. In this case, the strongly coupled hybrid modes arise from the fact that the electric field of SGPs experiences significant confinement in the $z$-direction due to the screening by the metal surface, effectively mimicking a plasmonic nanocavity~\cite{gan2012strong}. The diagram in Fig. \ref{Fig: coupling and loss intensity}(c) illustrates the coupling energy level splitting between the TMD phonons and the plasmonic nanocavity, i.e. the SGPs, which can be interpreted as two coupled harmonic oscillators~\cite{Hertzog2019,torma2014strong}. In this case, the frequencies of the hybrid modes will change along with the coupling strength, leading to anti-crossing in the dispersion. The plasmon-phonon interaction, represented by $\Omega$, defines the coupling strength~\cite{Novotny2010} and, at the minimum energy splitting, is calculated as~\cite{lavor2021tunable}
\begin{equation}
    \Omega =\frac{1}{2}[\omega_{+}(q)-\omega_{-}(q)]_{\text{min}}~.
    \label{eq: p-ph coupling}
\end{equation}
Here, $\omega_{+}(q)$ and $\omega_{-}(q)$ are the frequencies of upper and bottom hybrid modes as illustrated in Fig.~\ref{Fig: coupling and loss intensity}(b,c). For reference, the plasmon dispersion without the contribution of TMD phonons is represented by a solid purple line in Fig.~\ref{Fig: coupling and loss intensity}(b). In this case, to exclude the phonon contribution, the QEH calculation is performed using only the in-plane high-frequency dielectric constant of the individual layers at the optical limit, $\epsilon_{\parallel}^{\infty}$.

Finally, increasing the number of MoS$_2$ layers in the vdWhs increases the phonon density. Fig.~\ref{Fig: coupling and loss intensity}(d) illustrates the loss intensity calculated for the wave vector $q$ at the minimum energy splitting, where one observes that left and right peaks move further away as the number of layers increases, as a consequence of the effective increase in coupling strength due to the larger number of phonon modes.

\subsection{Strongly enhanced coupling due to a nearby metal surface}\label{sec. enhancing}

The interaction between SGPs and TMD phonon states can be categorized into three distinct coupling regimes: weak coupling (WC), strong coupling (SC), and ultrastrong coupling (USC). In order to determine the significance of the hybrid modes, it is important to compare their coupling strength ($\Omega$) with other key energy scales, for example, the phonon energy ($\hbar\omega_{ph}$) and the linewidth of the interacting system \cite{torma2014strong,lavor2021tunable}. Within the WC regime~\cite{cuartero2018light}, the interplay between phonon and SGPs dispersions is minimal, resulting in no significant changes to the dispersion profiles of either SGPs or phonons. In this case, $\Omega$ is sufficiently small to be neglected \cite{torma2014strong,Bitton2019}. In the SC regime~\cite{kockum2019ultrastrong,zhong2016non,orgiu2015conductivity}, mixed states can substantially modify plasmonic dynamics, leading to a pronounced anticrossing of the coupling modes in the $(q,\omega)-$plane, characterized by a vacuum Rabi splitting~\cite{santhosh2016vacuum,bitton2020vacuum}. The latter is defined as the frequency at which the probability amplitudes of two atomic energy levels fluctuate in an oscillating electromagnetic field. Moreover, when the interaction reaches sufficient strength, perturbations between the wave functions lead to alterations in the energy levels~\cite{vasista2020molecular,torma2014strong}. The USC is a distinct regime of electromagnetic interaction that enables a rich variety of intriguing physical phenomena. Within this regime, the plasmon-phonon coupling extends beyond merely altering the energy dispersion, and gains potential to modify the observable characteristics of the system. To quantify the coupling regimes, we normalize the coupling strength to its value at the phonon frequency of the minimum energy splitting, as $\eta = \Omega / \hbar\omega_{ph}$. Then, the WC regime is defined by $\eta<0.01$, the SC regime by $0.01\leq\eta<0.1$, and the USC regime by $\eta \geq 0.1$~\cite{kockum2019ultrastrong}.

\begin{figure}[!t]
\centering{}\includegraphics[width=\columnwidth]{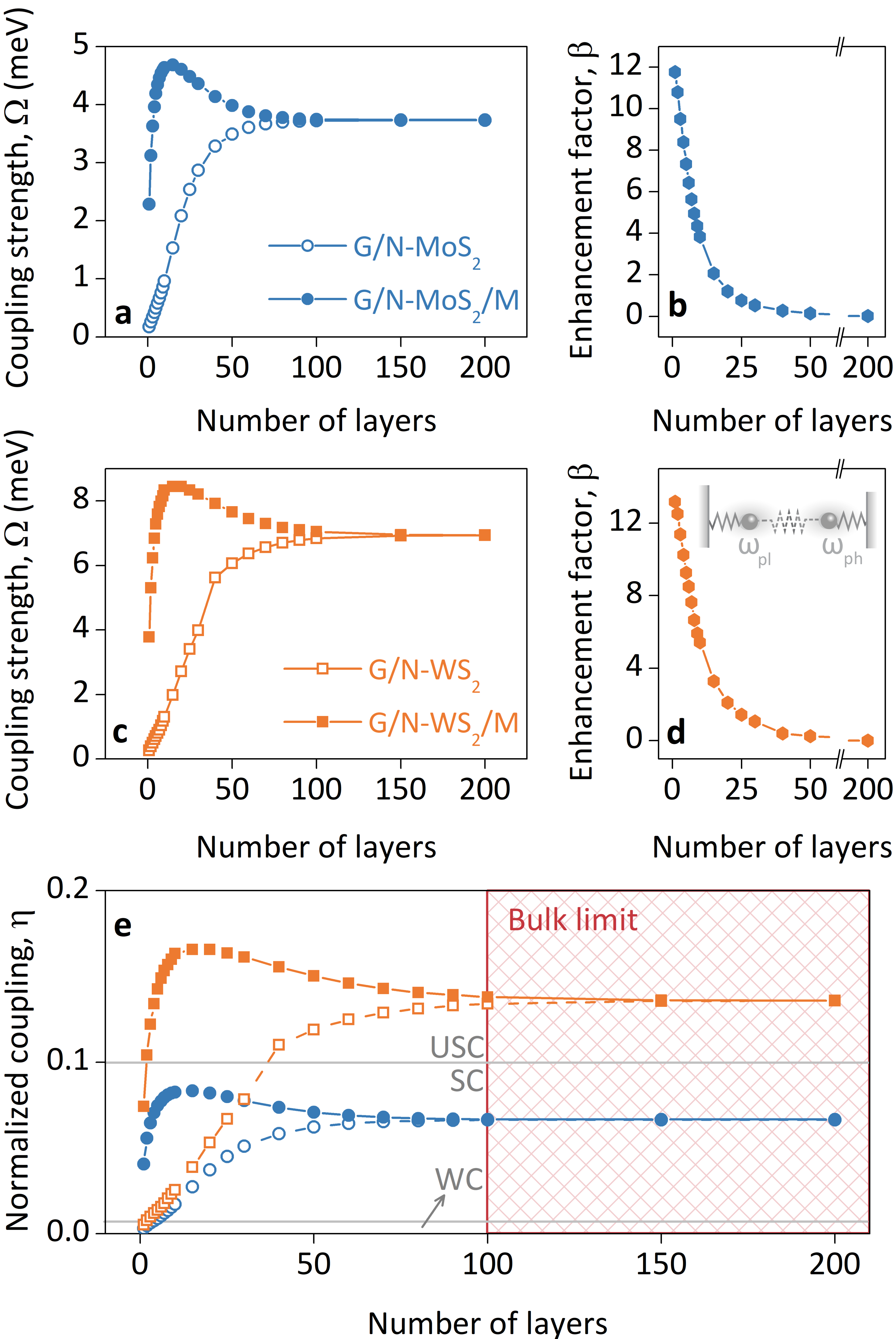}
\caption{(Color online) (a) SGPPPs coupling strength ($\Omega$) as a function of the number of TMD layers ($N$), assuming the graphene layer at Fermi energy $E_F=100~$meV on top of the N-MoS$_2$ (blue open symbols) and enhanced by a metal surface (blue closed symbols). (b) Enhancement factor $\beta$ due to the screening by the metal. Panels (c) and (d) show the same as (a) and (b), but for WS$_2$. (e) SGPPPs coupling strengths normalized to their respective monolayer phonon frequencies, defined as $\eta = \Omega / \hbar\omega_{A^{\prime\prime}_2}$. Three different regions, delimited by gray horizontal lines, represent the WC ($\eta < 0.01$), SC ($0.01 \leq \eta < 0.1$), and USC ($\eta\geq0.1$) regimes~\cite{kockum2019ultrastrong}. The hatched area represents the bulk limit of the SGPPPs coupling, reached for TMD thicker than $\approx100$ layers.}
\label{Fig3: Plasmon strength and bulk limit}
\end{figure}

Figure \ref{Fig3: Plasmon strength and bulk limit}(a) shows the coupling strength $\Omega$, i.e. the splitting of the SGPPPs as a function of the number $N$ of MoS$_2$ layers  in the absence (open symbols) and in the presence (filled symbols) of the metal. In both situations, the Fermi energy is $E_F=0.1~$eV. In the absence of the metal, the coupling strength in G/$N$-MoS$_2$ monotonically increases with $N$ until it reaches a maximum value, namely, its bulk limit value. In contrast, when considering the presence of the metal substrate, the coupling strength peaks at approximately ten layers of MoS$_2$, as observed in the results for G/$N$-MoS$_2$/M in Fig.~\ref{Fig3: Plasmon strength and bulk limit}. This suggests that, with the thickness of a monolayer of MoS$_2$ defined as $6.15~\text{\AA}$ in the QEH model, the effectiveness of screening provided by the metal surface starts to diminish for TMD thickness exceeding $\approx 60~\text{\AA}$. Beyond this thickness, the increased separation resulting from the increasing number of layers prevails the screening effect caused by the presence of the metal, thus leading to a decrease in the coupling strength, until it converges to its bulk value at large $N$. The enhancement of the coupling strength $\Omega$ driven by the metal surface for the case with several TMD layers (i.e. small $N$) is emphasized in Fig.~\ref{Fig3: Plasmon strength and bulk limit}(b), which presents the enhancement factor defined as $\beta = (\Omega_{\text{M}}-\Omega_{\text{WM}})/\Omega_{\text{M}}$, where $\Omega_\text{M}$ ($\Omega_\text{WM}$) is the coupling strength in the presence (absence) of the metal surface, as a function of the number of TMD layers.

Figures \ref{Fig3: Plasmon strength and bulk limit}(c) and (d) show the same quantities as panels (a) and (b), but for WS$_2$ instead of MoS$_2$. In this case, similar results are obtained, except that the coupling strength $\Omega$ is approximately 4 meV larger when compared to the results of MoS$_2$, even in the absence of a metal surface. We attribute this small difference to the fact that MoS$_2$ and WS$_2$ have different phonon masses. Namely, the plasmon-phonon coupling strength $\Omega$ can be understood by considering a simple system of two coupled classical harmonic oscillators~\cite{lavor2021tunable,torma2014strong}, as illustrated in the inset of Fig.~\ref{Fig3: Plasmon strength and bulk limit}(d), where $\Omega$ plays the role of the coupling energy between the masses in the oscillator~\cite{torma2014strong}. 

Figure ~\ref{Fig3: Plasmon strength and bulk limit}(e) shows the SGPPPs coupling energies normalized to the phonon frequencies $\omega_{\text{A}^{\prime\prime}_2}$ of a single TMD layer, i.e. $\eta = \Omega/ \hbar\omega_{\text{A}^{\prime\prime}_2}$, as a function of the number of layers $N$ in $N$-MoS$_2$ and $N$-WS$_2$. Three different regimes are delimited by gray horizontal lines: WC ($\eta < 0.01$), SC ($0.01 \leq \eta < 0.1$) and USC ($\eta\geq 0.1$)~\cite{kockum2019ultrastrong}. A notable result is achieved for the coupling between SGPs and the IR-active out-of-plane WS$_2$ phonon mode, which reaches the USC regime. When $N$ exceeds approximately 100 layers, the system attains bulk characteristics, leading to a convergence of both coupling scenarios (with and without the adjacent metal surface).

\subsection{The influence of doping on the SGPPPs spectra}\label{subsec: doping}

\begin{figure}[!t]
\centering{}\includegraphics[width=\columnwidth]{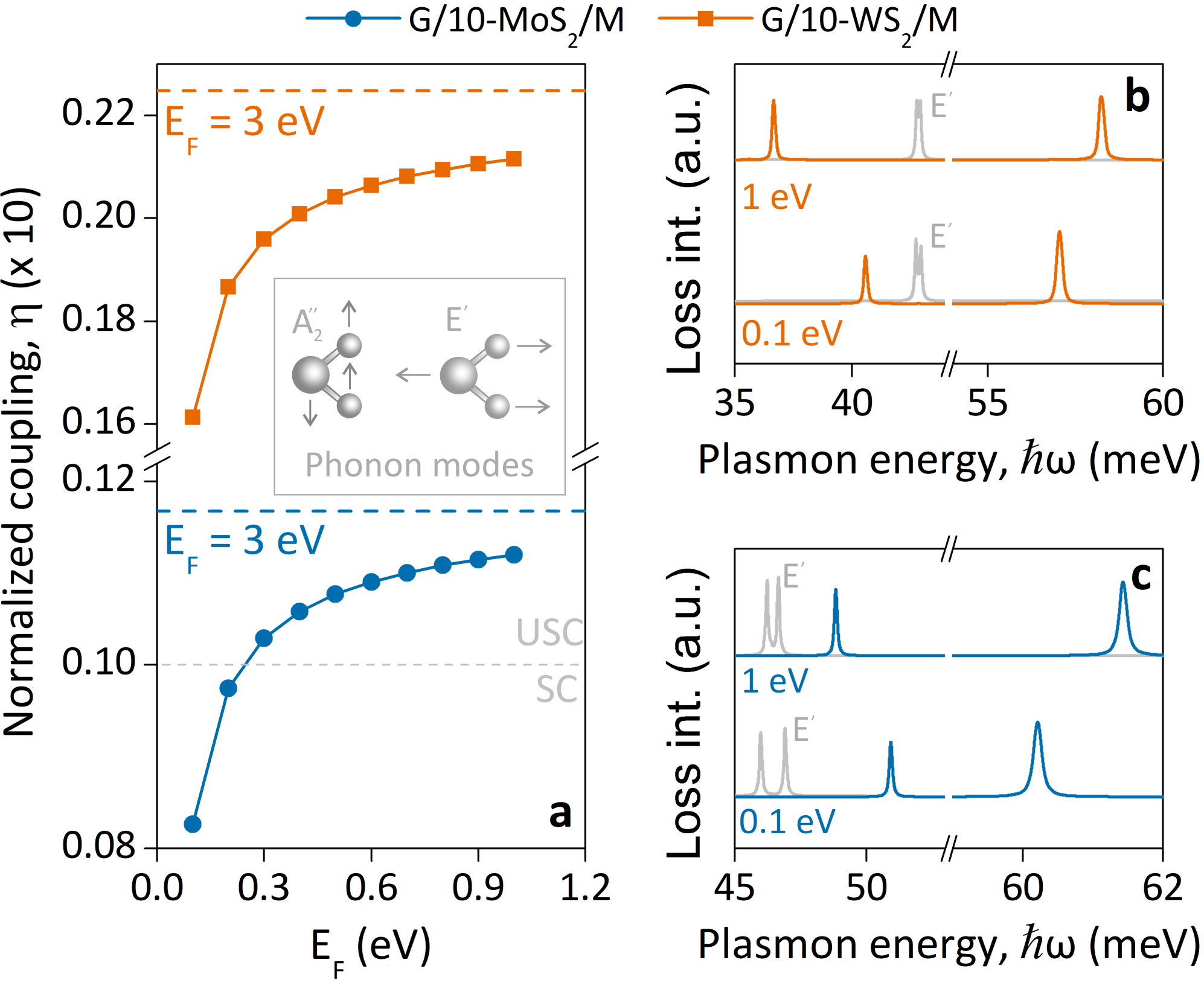}
\caption{(Color online) (a) The blue (orange) curve and symbols stand for the normalized coupling between SGPs and out-of-plane IR-active A$^{\prime\prime}_2$ MoS$_2$ (WS$_2$) phonon modes as a function of $E_F$. The horizontal dashed blue and orange curves are $\eta$ calculated for $E_F=3~$eV. The USC and SC regimes are delimited by a horizontal dashed gray line. The insets provide a visual depiction of the direction of vibration of the out-of-plane A$^{\prime\prime}_2$ and the in-plane E$^{\prime}$ TMDs phonon modes. Panel (b) and (c) are the loss intensity calculated at the MES for the SGPPPs (orange and blue curves), while the gray results are for the graphene plasmons and in-plane E$^\prime$ phonon mode.}
\label{Fig: doping}
\end{figure}

One of the most attractive features of graphene plasmonics lies in the possibility to manipulate and modulate nanoscale optical properties \textit{in situ}, by changing the graphene Fermi energy. This allows one to vary the carrier density within the graphene monolayer, thereby enabling the tuning of the plasmon wavelength ($\lambda_p$) to offset the reduction caused by the presence of the metal surface (cf. Fig.~\ref{Fig: electric field}). Furthermore, increasing the Fermi energy enhances the interaction of the electrons in graphene and their mirror charge due to the screening of the metal. To investigate this effect we consider G/10-MX$_2$/M vdWhs, with results shown in Fig.~\ref{Fig: doping}. The coupling between SGPs and out-of-plane IR-active A$^{\prime\prime}_2$  phonon modes is reinforced by the increased Fermi energy, similar to resonances in a photonic nanocavity~\cite{gan2012strong}. This situation is depicted in Fig.~\ref{Fig: doping}(a) by the solid blue (orange) line and symbols, where the normalized coupling between SGPs and out-of-plane IR-active A$^{\prime\prime}_2$ MoS$_2$ (WS$_2$) phonon modes increases with increased $E_F$. Note that the results for SGPs and out-of-plane IR-active A$^{\prime\prime}_2$ WS$_2$ phonon modes exclusively reside within the USC regime, i.e. it does not present a transition from the SC to USC as for MoS$_2$. The horizontal dashed blue and orange curvess are the normalized coupling at $E_F=3~$eV. Figure ~\ref{Fig: doping}(b) are the loss intensities calculated at the MES for the coupling between SGPs and out-of-plane A$_2^{\prime\prime}
$ phonon mode, represented by the orange lines, while the gray results are for the SGPs and in-plane E$^\prime$ phonon mode, for G/10-WS$_2$/M vdWhs. The results are for a Fermi energy at 0.1 and 1.0 eV. In fact, increasing $E_F$ enlarges the gap between the peaks of the out-of-plane coupled modes, thereby intensifying the coupling effect. Conversely, as illustrated by the solid gray lines in panels (b) and (c), the coupling between SGPs and in-plane IR-active E$^{\prime}$ phonon mode \textit{decreases} due to the increased confinement, resulting in a reduction of the in-plane electric field component (see Fig.~\ref{Fig: electric field}). Panel (c) is the same as in (b), but for MoS$_2$.

\subsection{Assessing vdWhs composition using SGPPPs}\label{sec: probing}

\begin{figure}[!t]
\centering{}\includegraphics[width=\columnwidth]{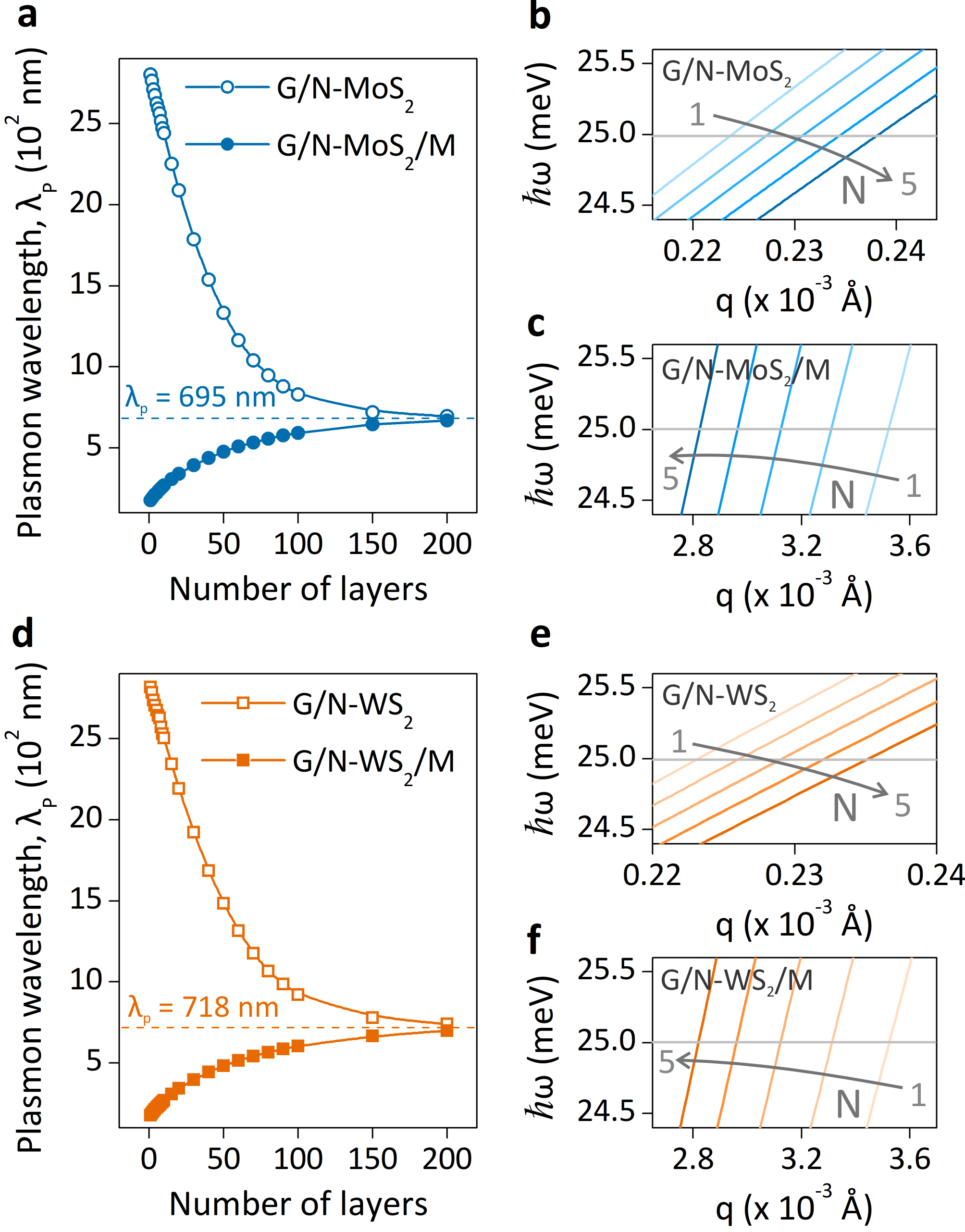}
\caption{(Color online) (a) The plasmon wavelength as a function of the number of MoS$_2$ layers in a G/$N$-MoS$_2$ vdWhs without a metal (open symbols) and with a metal as a substrate (solid symbols). The horizontal dashed blue line marks the plasmon wavelength ($\lambda_p = 695~$nm) for bulk MoS$_2$. Panels (b) and (c) show the plasmon dispersion for G/$N$-MoS$_2$ and G/$N$-MoS$_2$/M vdWhs, respectively. The gray arrow in these panels illustrates the increasing number of TMD layers, from $N=1$ to 5. The results for WS$_2$, corresponding to panels (a), (b) and (e), are shown in panels (d), (e) and (f), respectively.}
\label{Fig: apps wavelength}
\end{figure}

The increased sensitivity of acoustic graphene plasmons on the host platform presents an opportunity to explore the enhanced detection of the structure and composition of vdWhs, similarly as studied earlier for the optical plasmons~\cite{lavor2021tunable}. Figure \ref{Fig: apps wavelength} overviews the possible acoustic plasmon wavelengths for G/$N$-MX$_2$ and G/$N$-MX$_2$/M vdWhs, as a function of $N$. Without metal as a substrate, as illustrated by the blue open symbols in Fig.~\ref{Fig: apps wavelength}(a), increasing the number of layers leads to a reduction in the plasmon wavelength. Specifically, at a constant frequency of $\hbar \omega = 25~$meV ($\approx$ 6 THz), usual in s-SNOM experiments, adding more TMD layers leads to an increase in the plasmon wavevector $q$. This can be understood from the fact that the plasmon frequency in such vdWhs is inversely proportional to the square root of the environmental dielectric function, i.e. $\hbar \omega_{\text{pl}}\propto 1/\sqrt{\epsilon_{\text{env}}}$. Here, $\epsilon_{\text{env}}(q,d)$ depends on the wavevector $q$ and the thickness $d$ of the TMD, as discussed in Ref.~\cite{lavor2021tunable}. To emphasize this, the plasmon dispersion for G/$N$-MoS$_2$, from $N=1$ to 5, is presented in Fig.~\ref{Fig: apps wavelength}(b), and validates the enhanced screening effects from the TMD layers. Conversely, when a metal substrate is introduced, forming vdWhs such as G/$N$-MoS$_2$/M, the addition of TMD layers leads to diminished screening effects from the metal. This causes an increase in the plasmon wavelength with an increasing number of TMD layers, illustrated by the solid blue symbols in Fig.~\ref{Fig: apps wavelength}(a). This phenomenon is attributed to the dominant screening effects provided by the metal compared to those from the TMD. That is, adding more TMD layers, which act as spacers, effectively reduces the metal-induced screening effects. Consequently, this leads to a smaller wave vector $q$, as verified in Fig.~\ref{Fig: apps wavelength}(c), and, thereby, a larger plasmon wavelength ($\lambda = 2 \pi/q$). This analysis is also applied to WS$_2$, with similar findings displayed in Fig.~\ref{Fig: apps wavelength}(d)-(f). Therefore, while the presence of adjacent metal initially decreases the plasmon wavelength, it enhances its spatial resolution, i.e. the distinction in plasmon wavelengths for varied number of TMD layers becomes more pronounced, even for a vdWhs composed of two different TMDs, as will be discussed next.

\begin{figure}[!t]
\centering{}\includegraphics[width=1\columnwidth]{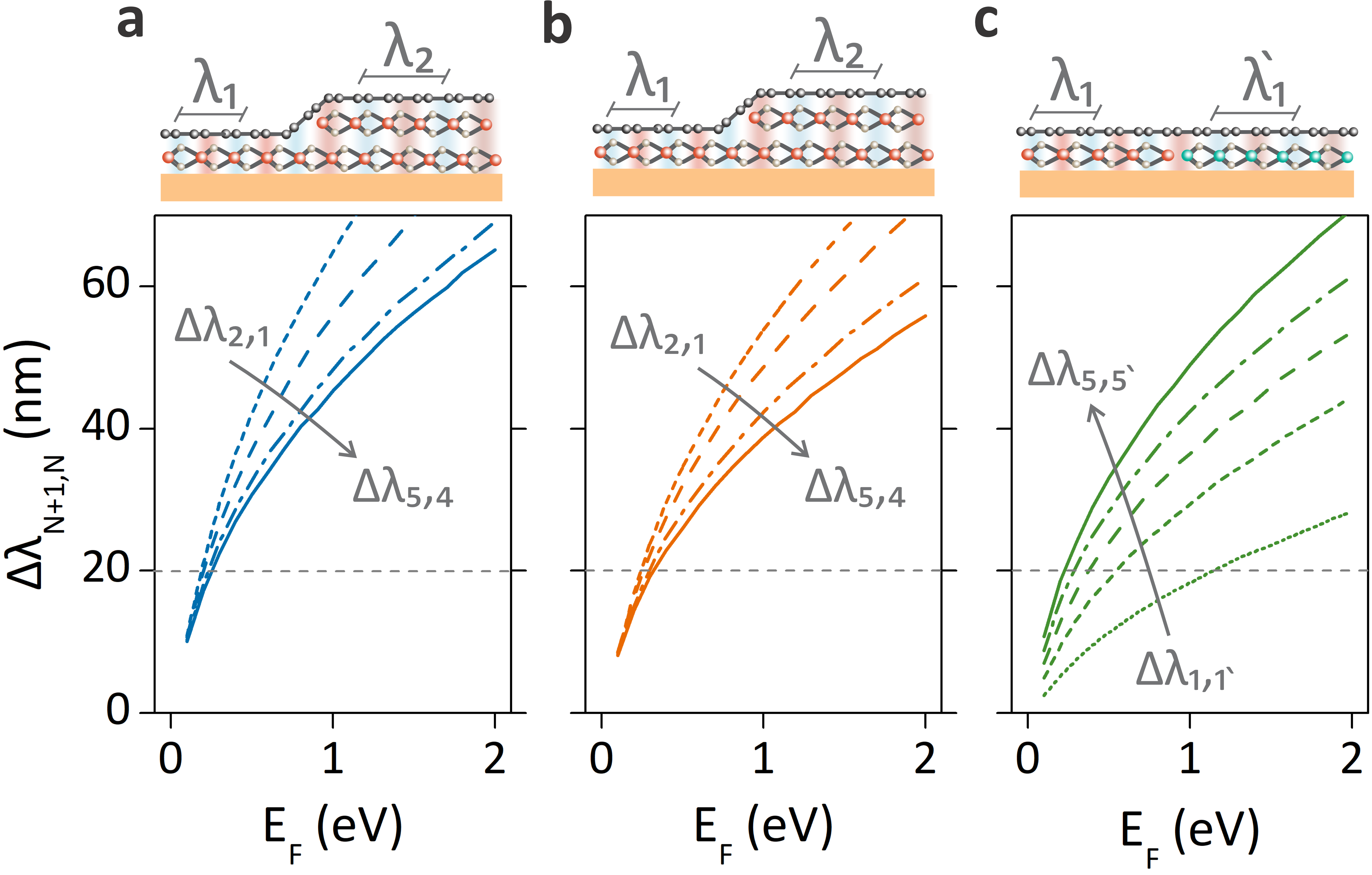}
\caption{(Color online) Plasmon wavelength resolution ($\Delta\lambda_{N+1,N}$) as a function of the Fermi energy, calculated as the difference between plasmon wavelengths found for $N+1$ versus $N$ TMD layers (for $N=1$ to 5) for (a) G/$N$-MoS$_2$/M and (b) G/$N$-WS$_2$/M vdWhs. Panel (c) presents the results for the same TMD thickness range of $N=1$ to 5, but for two different TMDs arranged in a side-by-side heterostructure. A horizontal dashed gray line marks the typical resolution threshold for s-SNOM experiments ($\approx20~$nm), indicating that structural detection with monolayer sensitivity is feasible based on our findings, even for a vdWhs composed of different laterally stitched TMDs, as in panel (c). The top sketches illustrate the vdWhs considered in each situation, with their respective plasmon wavelength.}
\label{Fig: wavelength as a function o EF}
\end{figure}

Figure \ref{Fig: wavelength as a function o EF} demonstrates that by adjusting the Fermi energy one can enhance the spatial resolution of the plasmon wavelength,defined here as the ability to distinguish between plasmon wavelengths for vdWhs with different numbers of TMD layers. To determine this resolution, we have calculated the difference between the plasmon wavelengths for vdWhs with $N$ and $N+1$ TMD layers at a given energy, i.e. as $\Delta\lambda_{N+1,N} = \lambda_{N+1} - \lambda_N $. Figures \ref{Fig: wavelength as a function o EF}(a) and \ref{Fig: wavelength as a function o EF}(b) show the results for vdWhs featuring a single TMD, for G/$N$-MoS$_2$/M and G/$N$-WS$_2$/M, respectively. The sketches above these panels illustrate the case of $N$ = 1 and $N$ = 2 stacked TMD layers, probed by a graphene layer laid on top of them. An even more remarkable result is found for the case of two different TMDs arranged in a side-by-side heterostructure, as illustrated by the sketch above Fig.~\ref{Fig: wavelength as a function o EF}(c). Results in this case show that the presence of the metal substrate, combined with the tunability of the Fermi energy in graphene, allows one to achieve single layer thickness resolution even for different materials beneath a graphene monolayer. In this case, the resolution is calculated as $\Delta\lambda_{N,N\prime} = \abs{\lambda_{N} - \lambda_{N\prime} }$, where $\lambda_{N}$ ($\lambda_{N\prime}$) is the plasmon wavelength considering a $N$-layer MoS$_2$ (WS$_2$) below the graphene.

\section{Conclusions}\label{sec:conclusion}

In summary, we have performed first principles-based calculations to investigate the effect of a proximal metal on the plasmonic dispersion of a monolayer graphene placed on top of a few-layer TMD, as well as on the coupling between the resulting screened graphene plasmons and the TMD phonons. 

The highly polarized electric field of the acoustic graphene plasmon that emerges due to the presence of the metal substrate leads to an amplified coupling strength with the out-of-plane A$_2^{\prime\prime}$ phonons of the TMD by up to one order of magnitude, thus reaching the ultra-strong plasmon-phonon coupling regime, which would be out of reach for this kind of mode in the absence of the metal. 

Our results also demonstrate the presence of a nearby metal increases the difference between plasmon wavelengths for graphene on different TMD materials, and/or on a different number of TMD layers in the heterostructure. Such marked improvement in the spatial resolution of plasmon wavelengths in real space allows for their use in probing the structural and compositional details of vdWhs down to the ultimate monolayer limit.  

Such significant role of metal substrates in boosting graphene plasmon resolution in real space opens up new avenues for engineering advanced plasmonic devices, especially in the field of ultrasensitive detection and high-resolution spectroscopy. Taken together, our results not only advance the understanding and control of the intricate interplay between plasmons and phonons within vdWhs, but also provide guidance for the exploration and realization of technological concepts based on such an interplay in presence of the metal-induced screening.

\section*{CRediT authorship contribution statement}
\textbf{I. R. Lavor}: Writing – original draft, Visualization, Validation,
Methodology, Investigation, Formal analysis, Data curation, Conceptualization \textbf{Z. H. Tao}:
Writing – review \& editing, Methodology, Investigation, Data curation. \textbf{H. M. Dong}: Writing – review \& editing. \textbf{A. Chaves}: Writing – review \& editing, Visualization, Supervision, Data curation, Validation. \textbf{F. M. Peeters}: Writing – review \& editing, Visualization, Validation. \textbf{M. V. Milo\v{s}evi\'c}: Writing – review \& editing, Visualization, Validation, Conceptualization, Supervision, Project administration, Investigation, Funding acquisition.

\section*{Data availability statement}
The data that support the findings of this study will be made available by the authors upon reasonable request.

\section*{Declaration of competing interest}

The authors declare that they have no known competing financial interests or personal relationships that could have appeared to influence the work reported in this paper.

\section*{Acknowledgement}
This work was supported by the Brazilian Council for Research (CNPq), through the UNIVERSAL and PQ programs, by the Brazilian National Council for the Improvement of Higher Education (CAPES), and by the Research Foundation - Flanders (FWO). Z. H. Tao gratefully acknowledges support from the China Scholarship Council (CSC).

\bibliographystyle{elsarticle-num}
\bibliography{ref}

\end{document}